\documentclass[onecolumn,12pt]{article}
\usepackage{graphicx,amsmath,amsfonts}

\DeclareMathAlphabet   {\mathsc}{OT1}{cmr}{m}{sc}

\def\[{\left [}
\def\]{\right ]}
\def\({\left (}
\def\){\right )}

\newcommand{\GeV}      {~\mathrm{GeV}}
\newcommand{\TeV}      {~\mathrm{TeV}}

\newcommand{\hc}       {\mathrm{\; h.c. \;}}

\newcommand{\Tr}{{\rm Tr}}

\newcommand{\gappeq}{\mathrel{\rlap {\raise.5ex\hbox{$>$}}
{\lower.5ex\hbox{$\sim$}}}}
\newcommand{\lappeq}{\mathrel{\rlap{\raise.5ex\hbox{$<$}}
{\lower.5ex\hbox{$\sim$}}}}

\newcommand{\lsim}{\,\raise.3ex\hbox{$<$\kern-.75em\lower1ex\hbox{$\sim$}}\,}
\newcommand{\gsim}{\,\raise.3ex\hbox{$>$\kern-.75em\lower1ex\hbox{$\sim$}}\,}
\newcommand{\LL}{\mathcal{L}}
\newcommand{\OO}{\mathcal{O}}
\newcommand{\half}{\frac{1}{2}}
\newcommand{\identity}{{\rlap{1} \hskip 1.6pt \hbox{1}}}
\newcommand{\nlsm}{nl$\sigma$m\,}
\newcommand{\nlsmns}{nl$\sigma$m}
\newcommand{\nn}{\nonumber}
\newcommand{\vth}{\vartheta_{\eta \phi}}

\begin{document}

\begin{titlepage}
\begin{flushright}
LBNL-52381\\
UCB-PTH-03/07\\
hep-ph/0306161\\
\end{flushright}
\vskip 2cm
\begin{center}
{\large\bf Scalar Dark Matter From Theory Space}
\vskip 1cm
{\normalsize
Andreas Birkedal-Hansen$^{(a)}$ and  Jay G. Wacker$^{(b)}$\\
\vskip 0.5cm

(a)
Department of Physics\\ 
University of California\\
 Berkeley, CA 94720 , USA\\
 and \\
Theoretical Physics Group\\ 
Lawrence Berkeley National Laboratory\\ 
Berkeley, CA 94720, USA

\vskip 0.3in

(b) Jefferson Physical Laboratory\\
Harvard University\\
Cambridge, MA 02138\\
\vskip .1in}
\end{center}

\vskip .5cm

\begin{abstract}
The scalar dark matter candidate in a prototypical theory space little Higgs model is investigated.  We review all details of the model pertinent to a relic density calculation. 
We perform a thermal relic density calculation including couplings to the gauge and Higgs sectors of the model.  We find two regions of parameter space
that give acceptable dark matter abundances.  The first region has a dark matter candidate with a mass $\mathcal{O}(100 \text{ GeV})$, the second region has a candidate with a mass greater than $\mathcal{O}(500 \text{ GeV})$.  The dark matter candidate in either region is an admixture of an $SU(2)$ triplet and an $SU(2)$ singlet, thereby constituting a possible WIMP (weakly interacting massive particle).

\end{abstract}

\end{titlepage}


\section{Introduction}

The ``little Higgs'' models provide a new mechanism to stabilize the 
weak scale wherein the Standard Model Higgs is a pseudo Goldstone boson~\cite{Arkani-Hamed:2001nc,Arkani-Hamed:2002pa,Arkani-Hamed:2002qx,Arkani-Hamed:2002qy}.  
The mass of the Higgs is protected by approximate non-linear global symmetries.
Subsets of these global symmetries are broken by couplings in the theory; 
however, this breaking occurs in such a way that any single 
coupling preserves sufficient symmetry to keep the Higgs massless.  
This ensures that there are no one loop quadratic divergences to
the Higgs mass and allows a two loop separation between the weak
scale and new physics.
The properties of these models allow for some distinct phenomenological 
signatures~\cite{Burdman:2002ns,Han:2003wu}.  

The dark matter problem is one that has long been optimistically tied to the 
theory of weak interactions.  At present time, it is not certain exactly what 
constitutes the dominant portion of the matter density in the universe, but 
its non-baryonic nature is generally agreed upon.  It happens that a stable 
neutral particle with Standard Model weak couplings and a weak scale mass 
gives roughly the correct amount of thermal dark matter.  Such an intriguing 
coincidence should not be overlooked in any theory attempting to explain the 
breaking of electroweak symmetry.  Since little Higgs (LH) models offer a 
mechanism to stabilize 
the weak scale, it is natural to ask if LH models have anything
to say about the dark matter problem.

In this paper we investigate the possibility that LH models may explain the 
observed abundance of dark matter.  We first review some general features of 
theory space LH models that are related to a dark matter analysis.  Then we specialize to the case of the $SO(5)$ theory space LH model contained in~\cite{Chang:2003un}.  Next we 
perform a 
relic abundance calculation including couplings of the dark matter candidate to the gauge and Higgs sectors of the model.  We find that a specific LH model allows proper dark matter 
abundances in two distinct interesting regions.  First, the correct relic abundance is achieved when the dark matter candidate, $N_1$, has mass $\mathcal{O}(100 \GeV)$.  A mass of $\mathcal{O}(100 \GeV)$ is allowed for either a weakly coupled or a 'super-weakly' coupled $N_1$.  By super-weak coupling we mean  that it does 
not couple significantly
to Standard Model (SM) weak vector bosons, it only couples to the TeV scale vector bosons.  
A second preferred regime exists where the dark matter candidate is heavy, with a mass greater than $\mathcal{O}(500 \GeV)$.  The preferred couplings of $N_1$ in this regime vary from weak (at $m_{N_1}$ of $\mathcal{O}(1 \TeV)$) to 'super-weak' (at $m_{N_1}$ of $\mathcal{O}(500 \GeV)$).  We discuss possible additional effects on the relic density calculation
(beyond couplings to the gauge and Higgs sectors) and find that they will usually only strengthen our conclusions.  

\section{\label{sec:LHbasics}Dark Matter Candidates in Little Higgs Models}

In terms of dark matter, LH models come in several varieties.  
Some LH models, referred to as `theory space' LH 
models~\cite{Arkani-Hamed:2001nc,Arkani-Hamed:2002pa,Arkani-Hamed:2002qx}, 
contain new possibly stable neutral particles.   Many other LH models, 
especially non-theory space models~\cite{Arkani-Hamed:2002qy}, have no potential dark matter candidates
associated with the little Higgs mechanism. 
For those theories that do contain the possibility of new stable neutral 
particles, one can place constraints on the LH model by requiring that this 
particle supply all of the necessary dark matter.  In the very least, one 
must require that this particle not provide {\it more} than the required 
dark matter.  If it provides less, then to explain the remaining dark matter
there must be other stable particles from the ultraviolet completion
to the LH model, potentially similar to particles in \cite{Nelson:2003aj}. 

At this stage it is useful to state what specific characteristics of LH models allow for a possible dark matter candidate and to compare these with models of broken supersymmetry.  In models of broken supersymmetry, it is usually R-parity that keeps the lightest supersymmetric particle (LSP) stable.  One way to define the charge of a given particle under R-parity is: $R=(-1)^{2j + 3B +L}$.  Here $j$ is the spin of the particle, $B$ is the baryon number of the particle and $L$ is the lepton number of the particle.  This results in every ordinary particle being even under R-parity and every supersymmetric partner being odd under R-parity.  If R-parity is a conserved quantity, the lightest R-odd particle must be stable.  This stability is one of the crucial ingredients that allow the LSP in models of broken supersymmetry to be a possible dark matter candidate.  Since the present number density of a thermal dark matter candidate is determined by the evolution of its abundance on a cosmological timescale, {\it any} sizable violation of R-parity will result in no observable present-day relic density.  The other necessary ingredient for dark matter is the correct combination of mass and annihilation cross section.  It is an intriguing, and much-studied, coincidence that a stable particle with a mass of order the weak scale and annhilation rate dominated by weak processes gives a present thermal relic abundance in the right range to explain the observed dark matter.  In models of broken supersymmetry, the lightest neutralino and sneutrino have just these additional characteristics, on top of one of them frequently being the LSP.

What characteristics of LH models might allow for the right dark matter abundance?  Are the requirements of stability, weak-scale masses, and weak coupling accurate and sufficient to provide the correct relic density?  We will find the necessity of weak-scale masses and weak coupling to be model-dependent.  However, let us repeat that the issue of stability is generally not a model dependent question.  Since cosmological timescales govern the annihilation (and possible decay) of a dark matter candidate, any such candidate with a lifetime shorter than the age of the universe will have completely decayed by today, resulting in no relic abundance.  Fortunately some LH models do contain symmetries that can protect the stability properties of some constituent fields.

Theory space LH models allow for a geometrical description of the group and field content.  Such a geometrical description can also be used to illustrate some of the symmetries of the models.  For instance, many theory space LH models contain a $\mathbb{Z}_4$ symmetry that can interchange the non-linear sigma model fields among themselves.  This symmetry is visualized in~\cite{Arkani-Hamed:2002pa} as rotations of $90^\circ$ on a square torus, though the symmetry exists in many theory space LH models without such a simple visualization.  For instance, this symmetry is contained in the model~\cite{Chang:2003un} we study in greater detail later on in this paper.  Because the lightest Higgs boson is contained in these non-linear sigma model fields, the $\mathbb{Z}_4$ symmetry is broken once electroweak symmetry is broken by the development of a Higgs vacuum expectation value (vev).  In the model we study in detail, the existence of the Higgs vev breaks this symmetry down to $\mathbb{Z}_2$.  It is the existence of this symmetry that allows theory space LH models to contain interesting dark matter candidates.  

Given the freedom contained in many LH models, it is possible that the $\mathbb{Z}_4$ and consequent $\mathbb{Z}_2$ symmetries be approximate instead of exact.  This is disastrous for the dark matter candidates we investigate, as has been explained above.  Therefore, our results apply only to models in which these symmetries can be made exact.  From here on we make the explicit assumption that at least the $\mathbb{Z}_2$ symmetry is exact.  As a result of our assumption, the lightest particle charged under this $\mathbb{Z}_2$ will be stable.  In theory space LH models, the Standard Model fields, including the Higgs, are even under this $\mathbb{Z}_2$.  In addition, theory space LH models always contain scalars $\phi$ and $\eta$, which are both odd under the $\mathbb{Z}_2$.  Thus, the lightest of these two will be stable and will constitute the model's only hope for a thermal dark matter candidate.  Interestingly enough, $\phi$ and $\eta$ are uncolored and also transform under the electroweak $SU(2)_W\times U(1)_Y$ as $\mathbf{3_0}$ and $\mathbf{1_0}$, respectively.  So only the $\phi$ can be considered a WIMP.  Regardless, there are important similarities between particle characteristics that exist in theory space LH models and characteristics that allow the LSP in broken supersymmetric theories to provide roughly the correct amount of dark matter.  

After electroweak symmetry breaking (EWSB), $\phi$ and $\eta$ decompose into 
\begin{eqnarray}
\phi \rightarrow ( \phi^0, \phi^\pm )
\hspace{0.5in} 
\eta \rightarrow \eta^0 .
\end{eqnarray}
Typically the $\phi^0$ and $\eta^0$ mix after EWSB due to
interactions of the form $\eta h^\dagger \phi h$.  We label the mass eigenstates $N_{1}$ and $N'_{1}$.  Electrically charged particles cannot serve as dark matter so one must require
$m_{\phi^\pm} > m_{N_1}$.
The $\phi$ and $\eta$ can either have the masses generated
completely radiatively or have a tree-level mass from an
``$\Omega$ plaquette'' \cite{Gregoire:2002ra}.
Schematically, the mass matrix is of the form:
\begin{eqnarray}
\LL_{\text{Mass}} &=& -\half (
m^2_{\phi\phi} (\phi^0)^2
+m^2_{\eta\eta} (\eta^0)^2
-2m^2_{\phi\eta} \phi^0 \eta^0)
- m^2_{\phi\phi} |\phi^\pm|^2
\end{eqnarray}
with
\begin{eqnarray}
\label{Eq:MassMatrix}
\nonumber
m^2_{\phi\phi} &=& m^2_G + m^2_S + m^2_E + m^2_\Omega\\
\nonumber
m^2_{\eta\eta} &=& m^2_S + m^2_E + m^2_\Omega\\
m^2_{\phi\eta} &=&  m^2_E 
\end{eqnarray}
where $m^2_G$,  $m^2_S$, $m^2_E$ and, $m^2_\Omega$ are
the masses from gauge interactions, scalar interactions, 
electroweak symmetry breaking, and possibly  a tree-level mass 
from an $\Omega$ plaquette.  Because $\eta$ is uncharged
under the Standard Model gauge interactions, it does not
receive a mass from this sector of the theory.   The gauge
contribution is typically the largest radiative contribution
to the masses because
\begin{eqnarray}
m^2_G \simeq \frac{3 g^2}{8 \pi^2} m^2_{W'} \log \(\frac{\Lambda^2}{m^2_{W'}}\)
\hspace{0.5in}
m^2_S \simeq \frac{\lambda}{8 \pi^2} m^2_{\phi'} \log \(\frac{\Lambda^2}{m^2_{\phi'}}\)
\hspace{0.5in}
m^2_E \simeq \frac{1}{4} \lambda v^2,
\end{eqnarray} 
with $m_{W'} \gsim 1.8 \TeV$ and $m_{\phi'} \simeq 1.5 \TeV$\footnote{The constraints on the masses of these particles come from precision electroweak constraints and limits on the breaking scale $f$~\cite{Chang:2003un,Csaki:2002qg,Hewett:2002px}.}.  Here $\phi'$ are the heavy scalars needed to cancel the quadratic divergences of the Higgs quartic coupling and $W'$ are the heavy gauge bosons needed to cancel the divergences coming from the Standard Model gauge bosons.  The parameter $\lambda$ determines the weight of the Higgs potential to be defined in Eqn.~\ref{eqn:higgspot}.  As usual, $v$ is the magnitude of the two Higgs vevs, $v^2 = v_{1}^2 + v_{2}^2$ for the two Higgs doublet model we study in detail in the next section.  The cut-off of the theory is $\Lambda \sim 4 \pi f \sim 10$ TeV and $g$ is the gauge coupling related to the TeV scale vector bosons.  The mass matrix for $\phi^0$ and $\eta^0$ can be diagonalized by an orthogonal transformation with a mixing angle, $\vth$.  If contributions from $\Omega$ plaquettes are ignored, the lightest eigenvalue of the mass matrix is of order $100 \GeV$.  Including $\Omega$ plaquettes can lift the lightest eigenvalue up into the $\TeV$ range.  

Depending on the full model, there can be a symmetry that interchanges $\phi^0 \leftrightarrow \eta^0$ in the absence of gauge interactions.  For the $SO(5)$ model we study in the next section, this symmetry just exchanges the $SU(2)_l$ and $SU(2)_r$ subgroups of the $SO(5)$.  This symmetry guarantees that the scalar ($m_S^2$) and EWSB ($m_E^2$) contributions between $\phi^0$ and $\eta^0$ exactly cancel in the calculation of this mixing angle, now given by

\begin{eqnarray}
\label{Eq: Mixing Angle}
\tan 2 \vth = \frac{2 m^2_E}{m^2_G} .
\end{eqnarray}
This angle becomes small when $m^2_G$ grows large.   In this
limit the dark matter candidate becomes predominantly 
$\eta^0$-like and does not have gauge interactions with the electroweak
vector bosons.  However, $\eta^0$ does have gauge interactions with the additional vector bosons of TeV-scale mass.  It is for
this reason we call this limit the ``super-weakly'' interacting
limit.  It is not uncommon to have $m^2_G$ large enough to yield $\cos^2 \vth \simeq 0.95$.

\subsection{The Minimal Moose}

The smallest theory space LH model is called the 
``Minimal Moose''~\cite{Arkani-Hamed:2002qx}.  The total global symmetry structure of the theory is $(G_L\times G_R)^4$.  This global symmetry structure is the essential feature that protects the Higgs from one loop divergences.  The theory contains four non-linear
sigma model (\nlsmns) fields $X_i$, $i=1,...,4$.  The presence of each \nlsm field breaks a global
$G_L\times G_R$ symmetry down to the diagonal group $G_D$.  Under global symmetry 
transformations $g_L$ and $g_R$, the \nlsm fields transform as
\begin{eqnarray}
X_i \rightarrow g_{L\,i} X_i g_{R\,i}^\dagger.
\end{eqnarray}
In some models there is also an exact $\mathbb{Z}_4$ discrete symmetry
that interchanges the \nlsm fields.
It is the $\mathbb{Z}_4$ symmetry that will result in a stable dark matter 
candidate.

Only a subgroup of the entire global symmetry group is gauged.
Inside $G_R^4$,  $SU(2)\times U(1)$ is gauged while inside $G_L^4$, 
a group $G$ is gauged which also contains an $SU(2)\times U(1)$ subgroup.   The four \nlsm fields break this extended gauge symmetry
down to the diagonal $SU(2)_W\times U(1)_Y$.   The
\nlsm fields transform under \begin{it}gauge\end{it} transformations
as bi-fundamentals:
\begin{eqnarray}
X_i \rightarrow g_{G}\, X_i\, g_{SU(2)\times U(1)}^\dagger.
\end{eqnarray} 
In terms 
of the vocabulary of 'moose' diagrams, this setup is a `two site - four link' 
model with each 'site' being a gauge group 
($G$ and $SU(2)\times U(1)$, respectively ),
and each `link' being a \nlsm field.  Since each link 
is charged under the gauge groups living at both sites, a geometrical picture 
can be drawn of the model as two dots connected by four lines.  This model is 
closely related to the $2\times 2$ toroidal moose 
of~\cite{Arkani-Hamed:2001nc}, which is a four site model and is also easy to 
visualize.  One views the $2\times 2$ toroidal moose as four dots, the 
corners of a square, connected by link fields to make the structure of a 
toroid.  The minimal moose is related to the $2 \times 2$ toroidal moose 
through orbifolding by a translation along each of 
the diagonals of the $2 \times 2$ toroidal moose.  

The gauge symmetries explicitly break 
the global symmetries.  The breaking of the $G_{L}^4$ global symmetry is 
accomplished by only the $G$ gauge 
transformations.  Similarly, the breaking of the $G_{R}^4$ global symmetry 
comes only from the $SU(2)\times U(1)$ gauge transformations.  Since the 
gauge structure of 
the model is known, one can write down the kinetic terms in the Lagrangian:
\begin{eqnarray}
\mathcal{L}_{K} = \frac{1}{2} \sum_{i} f^2 \Tr D_{\mu} X_{i} D^{\mu} X_{i}^{\dagger},
\end{eqnarray}
where $f$ is the 'pion decay constant' of the \nlsm.  The $D_{\mu}$ are the covariant derivatives, to be defined more explicitly below.

Up until this point we have kept the gauge group $G$ unspecified.  We have done this because
the minimal moose is, in some sense, modular.  Many general properties are determined irrespective of which group is chosen, as long as $G$
contains $SU(2)\times U(1)$.  However, each group
offers different predictions for TeV-scale physics.  As a result, some choices lead
to tension with measurements of precision electroweak observables.  For instance, in the original minimal moose~\cite{Arkani-Hamed:2002qx},
$G$ was chosen to be $SU(3)$.  The theory resulting from this choice can be significantly constrained by precision electroweak physics~\cite{Csaki:2002qg}.  In~\cite{Chang:2003un}
it was shown that $SO(5)$ (or equivalently $Sp(4)$) has a custodial
$SU(2)$ symmetry that allows a simple limit where precision electroweak
constraints are rather easily satisfied. 
There are several other groups that would work equally well such as
$SU(4)$ or $SO(7)$, but we choose here to only pursue the $SO(5)$ model described in~\cite{Chang:2003un}.
This group has the added bonus that there is a symmetry in the gaugeless limit that
interchanges $\phi^0$ and $\eta^0$ so that
Eq. \ref{Eq: Mixing Angle} is valid.  Another important motivation for $SO(5)$ is that
the approximate $\mathbb{Z}_4$ symmetry that is phenomenologically
necessary to prevent unacceptably large triplet vevs can, and will, be lifted
to an exact symmetry and will ensure the stability of the dark matter candidate.  

Even though the $SO(5)$ model is relatively unconstrained by precision electroweak physics, some constraints from such considerations do exist.  The most important comes from the breaking of custodial $SU(2)$ by electroweak symmetry breaking
vevs.  This breaking of custodial $SU(2)$ places a limit of $700 \GeV$ on the \nlsm breaking scale, $f$.  The custodial $SU(2)$ violation is
proportional to $\sin 2 \beta$, where $\tan \beta = v_2/v_1$ is the ratio of the vevs of the two Higgs present in the model (see below).  To suppress this effect, it is preferred for $\beta$ to be relatively small.  In
the $\mathbb{Z}_4$ symmetric limit that we are pursuing to find 
dark matter, there are no issues of triplet vevs because
there are no trilinear couplings between the Higgs doublets and
the $SU(2)_W$ triplets.    The effects upon the oblique $S$ and $T$
parameters from the new $\mathbb{Z}_4$ symmetric top sector would need to 
be calculated in this model.  However, it should be possible to make the effects small without significantly altering the limits.
In addition to the mildness of precision electroweak constraints, another benefit of this model is that it has a relatively minimal set of particles.

Let us explicitly summarize some details about the symmetries and symmetry breaking in the model we are investigating.  The $SO(5)$ minimal moose defined in \cite{Chang:2003un} has the expected site group structure $G\times G' = SO(5)\times [SU(2)\times U(1)]$.  The $[SO(5)]^8$ global symmetry is broken by the $SO(5)\times [SU(2)\times U(1)]$ gauge interactions.  The aforementioned custodial $SU(2)$ is approximate and comes about from the gauging of $SU(2)_{r}$ in addition to the secondary $SU(2)_{l}$, both from the $SO(5)$.  Thus, there exists a complete set of $SO(5)$ vector bosons with $SU(2)_{l} \times SU(2)_{r}$ representations $W^{l} \sim (\mathbf{3}_{l},\mathbf{1}_{r})$, $W^{r} \sim (\mathbf{1_{l},3}_{r})$, $V \sim (\mathbf{2}_{l},\mathbf{2}_{r})$.  Due to the embedding of $U(1)_{Y}$ inside the full $SO(5)$, the $W^{r}$ bosons split into $W^{r, \pm}$ and $W^{r,3}$.  $W^{r,3}$ is the field responsible for cancelling the quadratic divergence of the $U(1)_Y$ gauge boson.  If it is assumed that the $SO(5)$ gauge coupling is large, the Standard Model $W$ and $B$ are mainly composed of the $[SU(2)\times U(1)]$ gauge bosons.  The heavier orthogonal combinations, $W'$ and $B'$ are thus mainly composed of the $SO(5)$ gauge bosons.  The \nlsm\, fields $X_{i}$ break $SO(5)\times [SU(2)\times U(1)]$ down to the Standard Model $SU(2)_W\times U(1)_{Y}$, thereby giving $W'$ and $B'$ masses in the TeV range.  At scale of the $W'$ mass, the entire $\mathbb{Z}_4$ symmetry is still intact.  As the energy scale is lowered, the standard Higgs mechanism breaks electroweak symmetry and the Higgs vevs break $\mathbb{Z}_4\rightarrow \mathbb{Z}_2$.  However, as previously mentioned, this $\mathbb{Z}_2$ is sufficient to preserve the stability of the lightest eigenstate of the $(\eta,\phi)$ system.

In the case of $SO(5)$ the covariant derivative acting on the \nlsm fields 
is 
\begin{eqnarray}
D_{\mu} X_{i} = \partial_{\mu} X_{i} - i g_5 X_{i} T^{\[mn\]} W_{5, \mu}^{\[mn\]}+i\(g_{2} T^{l,a} W_{\mu}^{l,a}+g_{1} T^{r,3} W_{\mu}^{r,3}\)X_{i}.
\label{eqn:covder}
\end{eqnarray}
In this definition, $W_{5, \mu}^{\[mn\]}$ are the gauge bosons of the group $SO(5)$, $W_{\mu}^{l,a}$ are the $SU(2)$ gauge bosons and $W_{\mu}^{r,3}$ is the $U(1)$ gauge boson.   Expanding the \nlsm fields around $X_i = \identity$, one finds that the
kinetic term contains mass terms for the TeV-scale vector bosons
with 
\begin{eqnarray}
m^2_{W'} =  \frac{8 g^2 f^2}{ \sin^2 2 \theta},
\hspace{0.3in}
m^2_{B'} =  \frac{8 g'{}^2 f^2}{ \sin^2 2 \theta'}.
\end{eqnarray}
The mixing angles $\theta$ and $\theta'$ are defined as
\begin{eqnarray}
\tan \theta = \frac{g_2}{g_5},
\hspace{0.3in}
\tan \theta' = \frac{g_1}{g_5} .
\label{Eq: Gauge Mixing Angles}
\end{eqnarray}

At low energy the minimal moose has the same physics as any toroidal theory space~\cite{Arkani-Hamed:2001nc,Arkani-Hamed:2002pa}.  At tree level, two orthogonal 
combinations of pseudo-Goldstone boson multiplets are massless.  In terms of the linearized modes $X_{i} = \exp(i x_{i}/f)$, we take the massless combinations to be $x_{1}-x_{3}$ and $x_{2}-x_{4}$.  Continuing to follow the conventions of~\cite{Arkani-Hamed:2002qx}, these massless fields can be parametrized as

\begin{eqnarray}
X_{1} = X_{3}^{\dagger} \equiv e^{2i\(x+y\)/f}\\
X_{2} = X_{4}^{\dagger} \equiv e^{2i\(x-y\)/f}.
\end{eqnarray}
So in terms of the massless modes, there are two \nlsm\, fields,  $X=\exp(i x/f)$ and $Y = \exp(i y/f)$, called the little Higgses, that transform as adjoints under the diagonal global $SO(5)$ symmetry.  The low-energy Lagrangian in terms of these fields is:
\begin{eqnarray}
\label{Eq:Lagrangian1}
\LL_{LE} = 
\frac{f^2}{4} \Tr |\hat{D}_{\mu} X|^2 + 
\frac{f^2}{4} \Tr |\hat{D}_{\mu} Y|^2 +  \lambda f^4 \Tr XYX^\dagger Y^\dagger .
\end{eqnarray}
The covariant derivative, $\hat{D}$, acting on $X$ and $Y$, only contains the Standard Model $SU(2)_W\times U(1)_Y$ gauge fields.   $X$ and $Y$ also have interactions with the TeV-scale $W'$ and $B'$ vector bosons, but these have not been included in Eq.~\ref{Eq:Lagrangian1}.  For dark matter calculations, the most important interactions are given by Eq.~\ref{Eq:Lagrangian1} and
\begin{eqnarray}
\LL_{W' \text{ Int}}=  i g \cot 2 \theta \, \(\Tr W'_\mu X^\dagger \hat{D}^\mu X 
+\Tr W'_\mu Y^\dagger \hat{D}^\mu Y\) 
+\hc +\cdots\\
\LL_{B' \text{ Int}}= i g' \cot 2\theta' \, \(\Tr B'_\mu X^\dagger \hat{D}^\mu X 
+\Tr B'_\mu Y^\dagger \hat{D}^\mu Y \)
+\hc +\cdots,
\label{eq:heavyinteq}
\end{eqnarray}
where interactions involving more than one heavy vector boson have been suppressed.  The gauge mixing angles $\theta$ and $\theta'$ have been given in 
Eq. \ref{Eq: Gauge Mixing Angles}.  Bounds from non-oblique precision electroweak corrections require both $\theta$ and $\theta'$ to be small~\cite{Csaki:2002qg,Hewett:2002px,Chang:2003un,Gregoire:2003kr,Csaki:2003si}.  For simplicity we take the couplings of $X$ and $Y$ to $W'$ and $B'$ to be identical to the couplings to the Standard Model $W$ and $B$.

The potential in Eq.~\ref{Eq:Lagrangian1}, when expanded in terms of the linearized modes to leading 
order, is:
\begin{eqnarray}
\label{eqn:higgspot}
V(x,y) = \lambda \Tr [ x,y]^2 .
\end{eqnarray}
The linearized fields, $x$ and $y$, are adjoints under the unbroken
global $G_D$.  Under the electroweak $SU(2)_W\times U(1)_Y \subset G_D$,
the linearized fields decompose into several representations.  These representations include
$\phi$ and $\eta$, which are in the $\mathbf{3_0}$ and $\mathbf{1_0}$ 
representations of $SU(2)_W\times U(1)_{Y}$, respectively.  The decomposition also contains $h$, which is in the $\mathbf{2_{\half}}$ representation and is
identified as the Standard Model Higgs.  There can be additional
matter fields as well, but those are unimportant for our present discussion.

The discrete $\mathbb{Z}_4$ symmetry of the low energy theory in Eq.~\ref{Eq:Lagrangian1} takes
\begin{eqnarray}
X \rightarrow Y , \hspace{0.3in}
Y \rightarrow X^\dagger .
\label{eq:z4trans}
\end{eqnarray}
In~\cite{Arkani-Hamed:2002pa} this symmetry is understood as $90^\circ$ rotations on a square torus.  It is possible to lift this symmetry to the entire high energy theory and
also to make the symmetry exact, as is done in~\cite{Arkani-Hamed:2002pa}.  In many models there will be a transformation that takes
$X \rightarrow \Omega X \Omega$ and leaves the $\eta$ and $\phi$ fields invariant
but rotates the Higgs fields by $e^{i \pi}$.  For instance, such a transformation can take the form of $\Omega = \text{diag}(-1,-1,1)$ for $G = SU(3)$~\cite{Arkani-Hamed:2002pa} or $\Omega = \text{diag}(-1,-1,-1,-1,1)$ for our choice of $G = SO(5)$~\cite{Chang:2003un}.  For the model of~\cite{Chang:2003un}, this is just the $\mathbb{Z}_2$ symmetry from the $SU(2)_l$ subgroup of $SO(5)$, different from the $\mathbb{Z}_4$ symmetry of Eq.~\ref{eq:z4trans}.  After electroweak symmetry breaking the discrete $\mathbb{Z}_4\times \mathbb{Z}_2$ structure breaks leaving only an exact $\mathbb{Z}_2$ symmetry:
\begin{eqnarray}
X \rightarrow  \Omega X^\dagger \Omega 
\hspace{0.3in}
Y \rightarrow \Omega Y^\dagger \Omega .
\end{eqnarray}
Under this symmetry, the Standard Model fermions and Higgses are even, while $\phi\rightarrow - \phi$ and $\eta\rightarrow - \eta$.  This symmetry will leave the lightest triplet or singlet scalar stable and can result in dark matter.  It is important to note that to have a viable dark matter candidate it is necessary both that the $\mathbb{Z}_2$ be exact and that the dark matter candidate be the lightest field transforming under this $\mathbb{Z}_2$.  Generally the stable scalars in $X$ do not mix significantly with the stable scalars in $Y$.  This is because mixing is only generated at loop level through higher order interactions with the Higgs and the
top quark.  Each little Higgs,  $X$ and $Y$, contains a neutral $\eta$ and a neutral $\phi$, so there are four electrically neutral mass eigenstates that come from $\eta_{x,y}$ and
$\phi_{x,y}$.  The interactions of the $X$ and $Y$ fields are the same at tree level, so the minimal moose has two almost-identical copies of the
neutral $\eta$ and $\phi$ fields.  Thus we typically have two
nearly degenerate scalars $N_{1}$ and $N_{2}$ and the two heavier admixtures
$N_{1}'$ and $N_{2}'$.  Since $N_{1}$ and $N_{2}$ are nearly degenerate, the evolution of their relic abundances are intertwined.  Even though the heavier of the two (assumed to be $N_{2}$ for simplicity) will decay into the lighter one quickly (compared to the time scale involved in relic abundance evolution), both play important roles in the correct determination of the final dark matter density.  When we perform the relic density analysis we take this complication into account.

It is also possible to deform the minimal moose by adding $\Omega$ plaquettes.  At low energies these deformations result in terms being added to Eq. \ref{Eq:Lagrangian1}, of the form
\begin{eqnarray}
\LL_{\Omega\text{ Plaq}} = 
\kappa f^4 \Tr \Omega X \Omega X 
+\kappa f^4 \Tr \Omega Y \Omega Y +\hc.
\end{eqnarray}
These operators give tree-level masses for the scalars $\eta$ and $\phi$ while
leaving the Higgs massless.  They can be made to be symmetric under
the full $\mathbb{Z}_4 \times \mathbb{Z}_2$ high-energy discrete global symmetry.  Furthermore, these deformations allow us to lift the masses for the $\phi$s and $\eta$s from the weak scale to the TeV scale without affecting naturalness. 

It is important to remember that the mass matrix of the $(\eta,\phi)$ system is dominated by contributions from gauge interactions, and this frequently results in a lightest eigenstate that is mainly $\eta$-like.  Due to this strong dominance in determining the mass eigenstates, the gauge interactions also have a dominant influence on the relic abundance, as will be clearly shown in the next section.  For concreteness, we explicitly give the part of the $\phi$ covariant derivative containing $\phi_{x,y}^{0}$:

\begin{eqnarray}
\label{Eq:Dphi0}
\hat{D}_{\mu} \phi_{x,y} \supset \frac{1}{\sqrt{2}}\(\begin{array}{cc} \partial_{\mu} \phi_{x,y}^{0}  & i g W_{\mu}^{2} \phi_{x,y}^{0} - g W_{\mu}^{1} \phi_{x,y}^{0} \\ i g W_{\mu}^{2} \phi_{x,y}^{0} + g W_{\mu}^{1} \phi_{x,y}^{0} & -\partial_{\mu} \phi_{x,y}^{0} \end{array}\)
\end{eqnarray}

The charged and neutral $\phi$ fields also have interactions with the heavy vector bosons through interactions derived from Eq.~\ref{eq:heavyinteq}: 

\begin{eqnarray}
\label{Eq:Wpint}
\LL_{W' \text{ Int}}=  i g \cot 2 \theta \, \(\Tr W'_\mu \phi_{x}^\dagger \hat{D}^\mu \phi_{x}
+\Tr W'_\mu \phi_{y}^\dagger \hat{D}^\mu \phi_{y}\) 
+\hc +\cdots
\end{eqnarray}
\begin{eqnarray}
\label{Eq:Bpint}
\LL_{B' \text{ Int}}= i g' \cot 2\theta' \, \(\Tr B'_\mu \phi_{x}^\dagger \hat{D}^\mu \phi_{x}
+\Tr B'_\mu \phi_{y}^\dagger \hat{D}^\mu \phi_{y} \)
+\hc +\cdots.
\end{eqnarray}

We have left the $\phi$ in its $SU(2)$ triplet form in Eqs.~\ref{Eq:Wpint} and~\ref{Eq:Bpint}, but decomposition into $\phi^{\pm}$ and $\phi^0$ follows using the standard operations.  One can easily find the interactions involving $\phi^0$ by realizing that the neutral field comes only from the $a=3$ part of $\phi = \phi^a T^a$ where $T^a$ are the $SU(2)$ generators defined in the Appendix of Ref.~\cite{Chang:2003un}.  

The $\eta^0$ has no direct gauge interactions except for terms like $g'^2\eta^0 \eta^0 B'_{\mu} B'^{\mu}$.  These terms do not contribute to an extent that suppresses the relic density into the cosmologically preferred range. 

After the gauge interactions, the most important interactions come from the quartic potential.
To determine the quartic potential, one must define the vevs for the linearized modes $x$ and $y$ which contain the Higgses.  The directions of these vevs are:
\begin{eqnarray}
\langle x \rangle = \frac{v \cos\beta\cos \frac{\xi}{2}}{\sqrt{2}} T^{v\,0}
- \frac{v \cos \beta  \sin\frac{\xi}{2}}{\sqrt{2}} T^{v\,3}
\hspace{0.5in}
\langle y \rangle = \frac{v \sin \beta \cos \frac{\xi}{2}}{\sqrt{2}} T^{v\,0} 
+ \frac{v \sin \beta \sin\frac{\xi}{2}}{\sqrt{2}} T^{v\,3}
\end{eqnarray}
where we are again using the generators in the Appendix of Ref.~\cite{Chang:2003un}.
The quartic potential cannot stabilize EWSB when $\xi = 0, \pi$ 
because this is precisely where $[x,y]=0$.   In \cite{Chang:2003un}
the angle $\xi$ was taken to be $\frac{\pi}{2}$ to simplify the analysis, but it is possible that a symmetry can enforce this.   In this paper we use this
limit as a simplifying assumption.

Below we give the interaction potential for the $\(\eta,\phi\)$ system that we use in part to compute the annihilation rate in the next section.  This interaction potential can be derived directly from Eq.~\ref{eqn:higgspot}.  The $h^0$ is the usual lightest Higgs, $H^0$ is the heavy CP-even Higgs 
and $A^0$ is the CP-odd Higgs.  The $h^0-H^0$ mixing angle, $\alpha$,
is typically close to $\beta$ with this Higgs potential
although our results are generally insensitive to this.  Choosing $\alpha = \beta$ the potential is:
\begin{eqnarray}
\label{Eq:IntPotEtaPhi}
V =&& \frac{\lambda v}{8} h^0 (\cos^2 \beta (\phi_{x}^{0}-\eta_{x}^{0})^2+\sin^2 \beta (\phi_{y}^{0}-\eta_{y}^{0})^2)\nn\\
&+&\frac{\lambda v}{4} \sin 2 \beta \, H^0 ((\phi_{x}^{0}-\eta_{x}^{0})^2+
(\phi_y^0-\eta_y^0)^2)\nn\\
&+&\frac{\lambda v}{4} A^0 \, \eta_x^0 \eta_y^0
\end{eqnarray}
As can be seen in the above equation, it is only interactions with the CP-odd Higgs that allow the $N_{1}$ and $N_{2}$ eigenstates to coannihilate\footnote{Here we use the term {\it coannihilation} to refer to interactions between two {\it different} particle species in the thermal bath, such as $N_{1} N_{2} \rightarrow X$.  When we speak of standard {\it annihilation}, we mean interactions between two particles of the {\it same} species, such as $N_{1} N_{1} \rightarrow X$.}. 
Other than these interactions, the cosmological development of their relic densities is independent.

On the surface, the physical Higgs sector of this model resembles that in the MSSM.  However, one notable exception is the coupling of $b$ quarks
to the lightest Higgs.  In the MSSM, the coupling of the lightest Higgs to $b$ has a fixed $\tan \beta$ dependence.  In the minimal moose, there is
a freedom as to whether the lightest Higgs couples to $b$ quarks proportional to $\tan \beta$ or proportional to $\cot \beta$.  This can have a
non-negligible effect on the relic density.  In the analysis below, we choose the coupling to be proportional to $\tan \beta$, but we indicate how
the allowed LH parameter space would change had we chosen the coupling to be proportional to $\cot \beta$.

\section{\label{sec:dmbasics}Calculation of Relic Abundances}

Now we will use the general properties of the possible LH dark matter candidate
to calculate its relic abundance.  We will include all of the interactions listed in the previous section.  These would be expected to be the
dominant contributions.  Subdominant contributions might be expected from interactions with other heavy states in the theory.  We will find that a proper relic density generally prefers $N_{1}$ to have a larger mass than estimated in the literature on LH models,
though the notable exception exists of a super-weak eigenstate.

We start here by assuming that $N_{1,2}$ have some masses $m_{N_1}= m_{N_2}$ 
and can annihilate, as mentioned before, through a mixture of weak and 
super-weak gauge bosons and also the Higgs bosons\footnote{We remind the reader that we also include coannihilation effects since $N_{2}$ is assumed to be degenerate in mass with $N_{1}$.  These effects are explained in more detail in Section~\ref{sec:incco}.}.  We have given these interactions explicitly in Eqs.~\ref{Eq:Lagrangian1}, \ref{Eq:Dphi0}, \ref{Eq:Wpint}, \ref{Eq:Bpint} and~\ref{Eq:IntPotEtaPhi}.  From this we can calculate 
a thermally averaged cross section $\langle\sigma v \rangle$.  The thermally averaged cross section is used in the evolution equation for the number density of $N_{1}$, the Boltzmann equation:

\begin{eqnarray}
\frac{dn}{dt} = - 3 H n -\langle \sigma v \rangle \left[n^2-\left(n^{eq}\right)^2\right]
\end{eqnarray}

The Boltzmann equation shows that universal expansion always dilutes the number density of a particle species, but also that annihilations only become important once the particle species has diverted from equilibrium with the rest of the universe.  One would naturally expect this departure from thermal equilibrium to happen once the temperature of the universe has dropped below the mass of the particle, $T < m_{N_{1}}$.  Once the particle has decoupled from the surrounding thermal bath, the number density is still being diluted by Hubble expansion.  This expansion eventually shuts off the annihilation once the expansion rate, set by $H$, dominates over the annihilation rate, set by $\Gamma = n \langle\sigma v \rangle$.  After this event occurs, called 'freeze-out,' the number density of the particle species is essentially frozen except for further dilution from Hubble expansion.  Thus, one of the most important calculations to be done to determine a relic density is the calculation of the temperature at which freeze-out occurs, $T_F$.  This temperature is determined iteratively through its 
dimensionless inverse $x_F = m_{N_1}/T_F$:

\begin{equation}
x_F = \ln \left(\frac{0.038\, g\, m_{\text{Pl}} \,m_{N_1} \,
\langle \sigma v \rangle }{\sqrt{g_{*} \,x_F}}\right).
\end{equation}
Here $g$ is the number of degrees of freedom for the dark matter candidate, $m_{Pl}$ is the Planck mass, and $g_{*}$ is the number of effective relativistic degrees of freedom at the time of freeze-out of the dark matter particle.  For weakly interacting cold dark matter candidates, one normally finds $x_{F} \simeq 20$.  Having 
determined $\langle\sigma v \rangle$ and $x_{F}$, the most important formula 
we need is
\begin{equation}
\Omega_{dm} h^2 =\frac{1.07\times 10^{9}  \GeV^{-1}}{g_{*}^{1/2} \, 
m_{\text{Pl}}\, J\left(x_F\right)},
\label{eqn:Om1}
\end{equation}
where 
$J(x_F) = \int_{x_F}^{\infty} \langle\sigma v\rangle\, x^{-2} dx$.

\begin{figure}
\begin{center}
  \includegraphics[scale=1.]{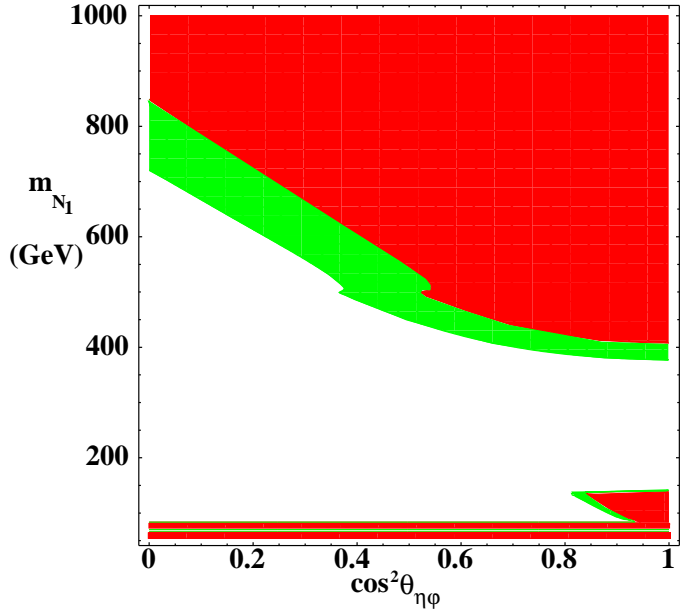}
\caption{\label{fig:lhdmplot1} 
{\bf Relic Density Contours for LH Models}. 
The preferred relic density regions are displayed for 
$\tan \beta=0.4$.  The green region illustrates a preferred value of 
$0.094 \leq \Omega_{dm} h^2 \leq 0.129 $ and the red regions are
regions where dark matter is over-produced, $\Omega_{dm} h^2 \geq 0.129$.}
\end{center}
\end{figure}

\begin{figure}
\begin{center}
  \includegraphics[scale=0.87]{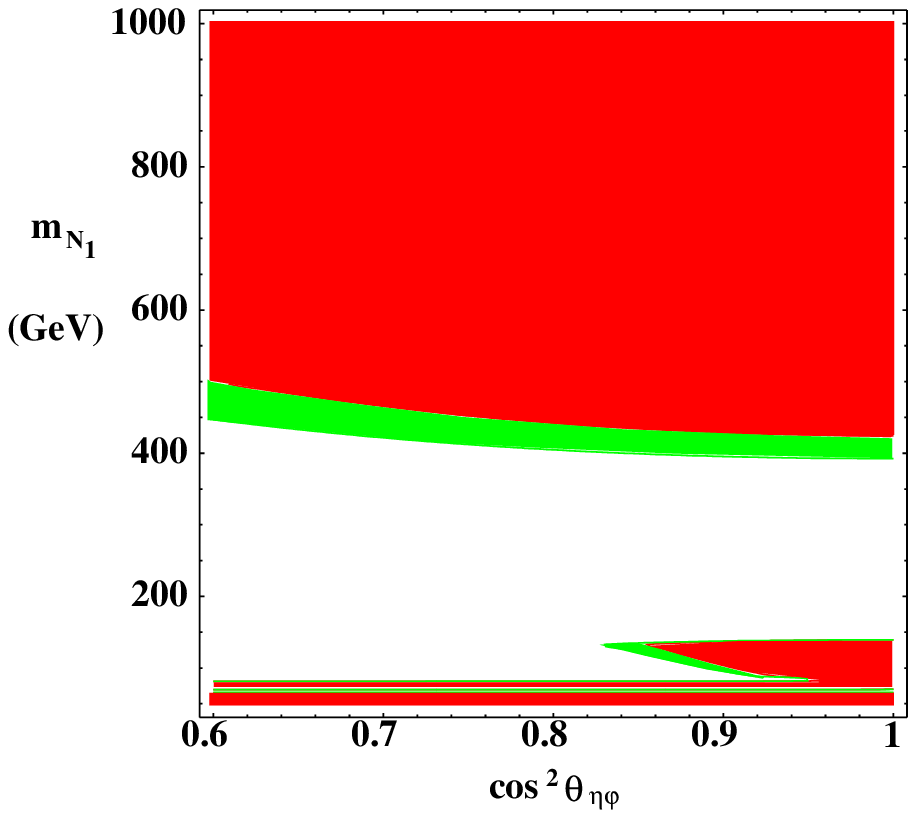}
\hspace{0.5in}
  \includegraphics[scale=0.87]{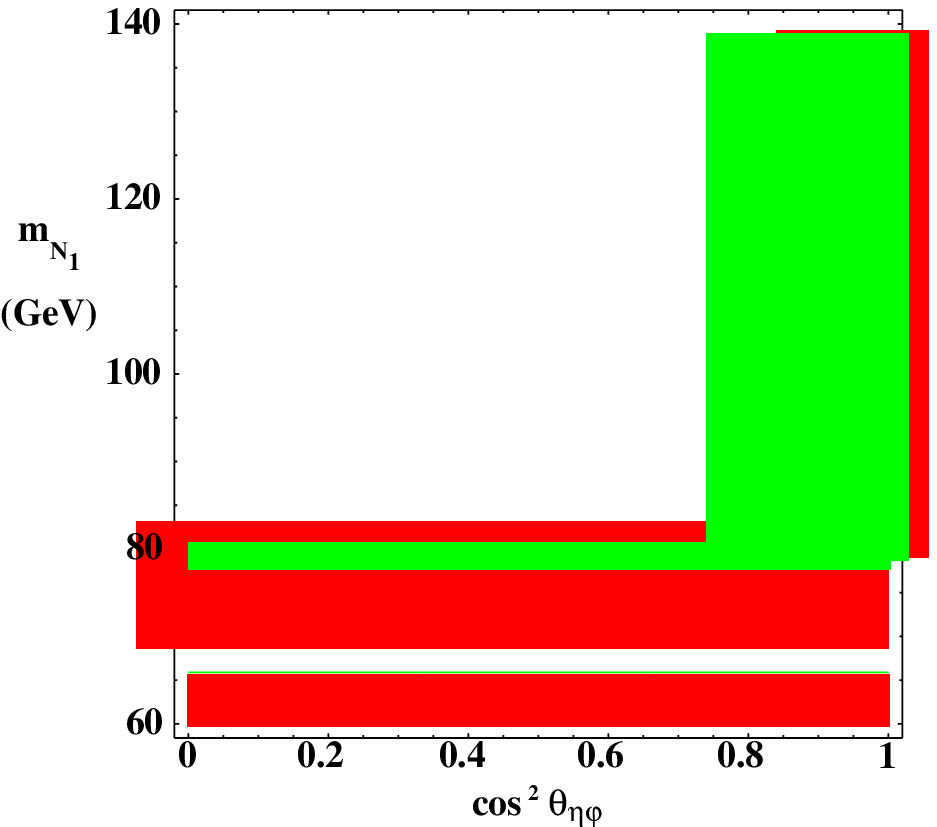}
  \caption{\label{fig:lhdmplot2} 
{\bf Relic Density Contours for LH Models}. 
This Figure is the same as Figure~\ref{fig:lhdmplot1} 
but magnified to show the finer detail in the low mass region.  The left plot also focuses in on the region with large $\cos^2 \vth$.}
\end{center}
\end{figure}

In many supersymmetric theories, one finds $m_{dm} \simeq 100 \GeV$.  
Recall that in supersymmetric theories, many of the annihilation channels of neutralinos are suppressed 
by a velocity factor of $(1-4 m_{dm}^{2}/s)$ because the initial state particles are fermions.  Here $s$ is the squared center-of-mass energy.  This suppression is even more severe in neutralinos because they are Majorana fermions, but the effect exists in some annihilation channels of Dirac fermions as well.  Because the dark matter candidates are often scalars in theory space LH models, this suppression factor is absent.  Therefore, the annihilation 
is more efficient and higher masses are required to achieve the correct relic 
density than in models involving neutralino dark matter.  This absence of velocity suppression is also important in other models with scalar dark matter candidates~\cite{Frieman:1991,McDonald:1993ex,Bertolami:1999sj,Peebles:2000yy,Goodman:2000tg,Matos:2000xi,Bento:2000vu,Bento:2001yk} such as inelastic sneutrino dark matter~\cite{Smith:2001hy}.  The heavier required masses can clearly be seen in Figure~\ref{fig:lhdmplot1} where
contours of constant relic density are plotted in the 
$(\cos^2 \vth, m_{N_{1}})$ plane.  Here $\cos^2 \vth$ is the mixing 
parameter defined in the discussion leading up to Eq.~\ref{Eq: Mixing Angle}.  It is the mixing parameter that determines the admixture of weak and super-weak strength with 
which $N_{1}$ can annihilate.  In this and following plots, the green (light shaded) 
regions are cosmologically preferred with 
$0.094 \leq \Omega_{dm} h^2 \leq 0.129 $~\cite{Spergel:2003cb}.  The red (dark shaded) regions are cosmologically 
excluded, having a relic density $\Omega_{dm} h^2 \geq 0.129 $.  In Figs.~\ref{fig:lhdmplot1} and~\ref{fig:lhdmplot2} we have chosen $\tan \beta = 0.4$, a moderate value that is still 
allowed by precision electroweak constraints \cite{Chang:2003un}.  We have 
also fixed the lightest Higgs mass to be $m_{h^{0}} = 136 \GeV$, the heavy Higgs 
masses to be $m_{H^0}= m_{H^\pm} = m_{A} = 500 \GeV$, and the 
heavy super-weak gauge boson masses to be $m_{W'} =m_{B'} = 1800 \GeV$.  Furthermore, here 
we take the Higgs boson to couple to the $b$ quark proportional to 
$\tan \beta$.  
As mentioned previously, this is an arbitrary choice.  We explain below what 
happens if this choice is instead changed to $\cot \beta$.

In Figure~\ref{fig:lhdmplot1} we can see that requiring the LH model to supply all required dark matter allows 
$m_{N_{1}}\sim 100 \GeV$ for any value of $\cos^2 \vth$, thus signalling a robust weak-scale dark matter candidate.  However, it is important to note that $\cos^2 \vth \geq 0.8$ is required to have $m_{N_1}$ not be constrained to lie on $m_{h^{0}} /2 \pm \delta$ or $m_W$, where $\delta$ is roughly $3 \GeV$.  This behavior will be explained in the next paragraph.  Between $m_W$ and $m_{h^{0}}$, a preferred region exists with the aforementioned $\cos^2 \vth \geq 0.8$.  Above $m_{h^{0}}$, annihilation to an $h h$ final state again reduces the relic density beyond relevancy.  Once above $m_{h^{0}}$, one must go beyond $m_{N_{1}} \sim 500 \GeV$ in order to find cosmologically 
preferred regions.  Why are masses around 
$400 \GeV$ required above $m_{h^{0}}$ for a general mixed $N_{1,2}$?  The single most important 
influence causing this characteristic is the aforementioned lack of 
suppression in the thermally averaged cross section, due to $N_{1,2}$ being 
scalars.

First, below $m_W$, there are no efficient annihilation channels to deplete $N_1$ except for resonant annihilation through the lightest Higgs boson.  As this resonance is approached from either above or below, the annihilation cross section rises rapidly until the dark matter constraint is satisfied for a brief range in $m_{N_1}$.  This situation is different from other paradigms such as supersymmetry because $N_1$ has no direct coupling to the neutral gauge bosons, so there can be no s-channel annihilation through $Z$ or $\gamma$.  Second, once $m_{N_1} > m_W$, efficient annihilation occurs to $W^+ W^-$.  Such significant annihilation depresses the relic density below preferred values except for the high-$\cos^2 \vth$ region.  It is this high-$\cos^2 \vth$ region where the $N_1$ is primarily an $SU(2)$ singlet, so it cannot annihilate effectively to $W^+ W^-$.  Above $m_W$, the preferred region gradually starts to prefer lower values of $\cos^2 \vth$ as $m_{N_1}$ increases.  This is because $\Omega_{DM} h^2 \sim m / \langle\sigma v\rangle$, so in order to keep $\Omega_{DM} h^2$ constant for a higher mass, the thermally averaged annihilation cross section must also increase. This trend continues until $m_{N_1} = m_{h^{0}}$. 

As was mentioned above, a region below $200 \GeV$ exists with the right relic abundance.  This region is below the $t \bar{t}$ and 
$h^{0} h^{0}$ production thresholds, so annihilation is not extremely efficient barring resonances.  We display this low-$m_{N_{1}}$ region in greater detail 
in the right plot of Figure~\ref{fig:lhdmplot2}.  Below $m_W$, there are no efficient annihilation channels to deplete $N_1$ except for resonant annihilation through the lightest Higgs boson.  As this resonance is approached from either above or below, the annihilation cross section rises rapidly until the dark matter constraint is satisfied for a brief range in $m_{N_1}$.  This situation is different from other paradigms such as supersymmetry because $N_1$ has no direct coupling to the neutral gauge bosons, so there can be no s-channel annihilation through $Z$ or $\gamma$.  This can be seen from the explicit expression for the covariant derivative, Eq.~\ref{Eq:Dphi0}.  Second, once $m_{N_1} > m_W$, efficient annihilation occurs to $W^+ W^-$.  Such significant annihilation depresses the relic density below preferred values except for the high-$\cos^2 \vth$ region.  It is this high-$\cos^2 \vth$ region where the $N_1$ is primarily an $SU(2)$ singlet, so it cannot annihilate effectively to $W^+ W^-$.  Above $m_W$, the preferred region gradually starts to move towards lower values of $\cos^2 \vth$ as $m_{N_1}$ increases.  This is because $\Omega_{DM} h^2 \sim m / \langle\sigma v\rangle$, so in order to keep $\Omega_{DM} h^2$ constant for a higher mass, the thermally averaged annihilation cross section must also increase. The trend continues until $m_{N_1} = m_{h^{0}}$. Finally, at $m_{h^{0}}$, the $h^{0} h^{0}$ production 
threshold is crossed and the relic density is uniformly reduced below a 
relevant level.  If one can imagine this parameter space absent of the 
thresholds and resonances, a general trend towards heavier masses at lower 
$\cos \vth$ is apparent if $m_{N_1} > m_W$.  This continues for the entire parameter space 
and is due to the strong $\cos \vth$ dependence in the couplings of $N_{1,2}$ to 
the Standard Model $W^{\pm}$.

Between $200$ and $400 \GeV$ in Figure~\ref{fig:lhdmplot1} a region is visible 
without a significant relic density.  This is due primarily to efficient annihilation into $t \bar{t}$ and $h^{0} h^{0}$ final states.  Preferred regions
arise again above $400 \GeV$ and move towards the weak end of the mixed region (low $\cos^2 \vth$) by $800 \GeV$.  This region between $400$ and $800 \GeV$ is dominated by annihilations to
$W^{+} W^{-}$, so the $\cos \vth$ dependence comes directly from mixing angle effects on the gauge couplings of $N_{1,2}$.  The main
additional structure in this region is the slight kink at $500 \GeV$ which is due to the production threshold for heavy Higgs final states.  The region above $1000 \GeV$ is excluded experimentally because 
annihilation is not efficient enough to reduce the relic density 
appropriately.   It is possible that UV completions to this model may provide 
additional annihilation channels.  
However, the mass of the dark matter particle eventually becomes limited by partial-wave
unitarity considerations and one has a difficult time avoiding the upper dark 
matter bound for a weakly coupled thermal relic of mass above $2 \TeV$~\cite{Griest:1989wd,Jungman:1995df}.

How sensitive are these regions of parameter space to the assumptions we have 
made?  The most robust regions are the super-weak area between 
$m_{N_{1}}\simeq 80 \GeV$ and $m_{N_{1}}\simeq 140 \GeV$ and the sliver of parameter space below $m_{h^{0}}/2$, shown in detail in the right plot in
Figure~\ref{fig:lhdmplot2}.  If one assumes that the $b$ quark couples 
proportional to $\cot \beta$ instead of $\tan \beta$, then the entire region 
between $m_{h^{0}}/2$ and $m_W$ becomes completely depleted of relevant dark matter, but the 
region between $m_{N_{1}}\simeq 80 \GeV$ and $m_{N_{1}}\simeq 140 \GeV$ 
remains unscathed and the $\cos^2 \vth$-independent sliver below $m_{h^{0}}/2$ remains, but drops down to around $50 \GeV$.  Additionally, if one takes a very small 
value for $\tan \beta$, say around $0.1$, then the preferred band in the region 
between $400 \GeV$ and $1000 \GeV$ in Figure~\ref{fig:lhdmplot1} gets pushed up to values between
$1700 \GeV$ and $2000 \GeV$, though the existence of the band is robust.  
So, regardless of the assumptions that are made, regions with an acceptable 
relic density exist below $m_{h^{0}}/2$ and between $80$ and $140\GeV$ for large values of 
$\cos^2 \vth$ and also above $900 \GeV$ for any value of $\cos^2 \vth$.

Dark matter direct detection limits depend on the mass of the relic 
particle.  Direct detection experiments are designed to best detect particles 
roughly in the $m_{N_1}= 80 - 140 \GeV$ range\footnote{
Direct detection possibilities for the super-weakly coupled dark matter
candidate are suppressed compared to typical supersymmetric candidates for two important reasons.
First, there are no intermediate colored states that couple the candidate
to nuclear matter.  Second, there are no couplings between $N_{1,2}$ and the neutral weak vector bosons.  Thus, nuclear scattering will occur dominantly through s-channel Higgs bosons, which is suppressed by Yukawa couplings.}.  In this model, a mass in this range requires a high value of 
$\cos^2 \vth$.   To achieve this, it requires a large gauge contribution
to the $\phi$ mass which translates into heavier super-weak vector bosons.    
This is the limit that precision electroweak observables prefer \cite{Chang:2003un}.  A study of direct and indirect detection prospects is beyond the scope of this paper, but is the subject of current ongoing work~\cite{Us4}.

Since gauge contributions dominate the mass matrix (Eq.~\ref{Eq:MassMatrix}) for the neutral $(\eta,\phi)$ system, Eq.~\ref{Eq: Mixing Angle} implies that large $\cos^2 \vth$ happens easily.  To investigate this region in more detail, we have focused on 
large $\cos^2 \vth$ in the left plot in Figure~\ref{fig:lhdmplot2}.  Taking this plot to represent the 
`natural' portion of the full parameter space, we can see that over half of 
the region (the part with red coloring) is ruled out on cosmological grounds.  Most of the rest of the 
parameter space has a relic density too low to be cosmologically relevant.  
The regions with acceptable relic density can be grouped into two 
categories.  The first region occurs between $400 \GeV$ and $900 \GeV$
and is due mainly to final states $W^{+} W^{-}$, $t \bar{t}$ 
and $h^{0} h^{0}$.  This region does not require any tight relationship between $N_{1,2}$ 
and another particle.  The second category is the low-$m_{N_{1}}$ region below 
$m_{N_{1}}=200 \GeV$.  If, for example, we take $\cos^2 \vth = 0.95$ and review Figure~\ref{fig:lhdmplot2}, 
we find that the dark matter candidate must take very specific mass values: 
$m_{N_{1}} = m_{h^{0}}/2 \pm \delta, m_{W}, m_{h^{0}}$, or $ 400 \pm 50 \GeV$.  
Again, $\delta$ is a small number around $3 \GeV$.  While it is possible for 
$\cos^2 \vth$ to be large without being around $0.95$, the predictive power of 
such a scenario in terms of dark matter is intriguing.

In summary, we find that LH models with a general mixed $N_{1}$ state do not provide 
enough relic density to explain dark matter unless either additional 
$\Omega$ plaquettes are included to lift $m_{N_{1}}$, or $m_{N_1}$ is chosen to have very specific values, namely $m_W$ or $m_{h^{0}}/2 \pm 3 \GeV$.  However, we do find 
that the most likely scenario, that of a super-weakly interacting $N_{1}$ with 
$m_{N_{1}}\sim 100 \GeV$,  is able to provide for the correct relic density, although the preferred region is quite narrow.

\subsection{\label{sec:incco}Other Possible Effects}

At this point it is natural to inquire about other contributions to the relic density calculations that might change our results.  As we will describe, additional effects will
likely only further reduce the available parameter space.  Possible contributions fall into one of three categories: additional annihilation
channels, coannihilation channels, and entropy generation after $N_{1}$ freeze-out.  We will discuss each of these in turn.

There are almost certainly other ways that two $N_{1}$ particles can 
annihilate.  These new channels may come from other loop-suppressed 
operators, or possibly from operators that appear in other specific models.  
Any time new annihilation channels are allowed, one expects the thermally 
averaged cross section to be increased.  This will further suppress the relic 
density of $N_{1}$.  We have already seen this in 
Figures~\ref{fig:lhdmplot1} and \ref{fig:lhdmplot2}
when the resonant annihilation and particle production thresholds were 
crossed.

As mentioned previously, it is also possible for the mass of $\phi_{1}^{+}$ (the lighter of $\phi_{x}^{+}$ and $\phi_{y}^{+}$) to be degenerate with the mass of $N_{1}$ to within 5-10\%.  Such a scenario can occur when the mass contributions coming from the $\Omega$-plaquettes dominate the mass matrices.  While this situation is unlikely for light $N_{1}$s, it is not unnatural for heavy $N_{1}$s, whose masses can be prinicipally determined by $\Omega$-plaquettes.  When this degeneracy happens, the two particles freeze out of equilibrium at about the same time, and the interactions of $\phi_{1}^{+}$ can
significantly affect the resulting relic density of $N_{1}$~\cite{Binetruy:1983jf,Griest:1990kh}.  We can make the normal assumption that
$\phi_{1}^{+}$ decays into $N_{1}$ on a time scale much shorter than the Hubble expansion time scale.  Thus, instead of treating the evolution of
the number density $n_{i}$ of each species separately, we can evolve the {\it total} number density $n=n_{N_{1}}+n_{\phi_{1}^{+}}$.  To do this, we
need to approximate the thermally averaged cross section as

\begin{equation}
\langle \sigma v \rangle = \sum_{i,j} \sigma_{ij} \frac{n_{i}^{eq}}{n^{eq}}\frac{n_{j}^{eq}}{n^{eq}},
\end{equation}
where the $eq$ superscript denotes values at equilibrium and $i$ and $j$ denote particle species to be summed over ($N_{1}$ and $\phi_{1}^{+}$ for this example).  From this formula we can see that just adding a coannihilator with equal annihilation
cross sections to the original particle has little effect on the relic density.  In fact, this is the case for our analysis already.  We have
included the fact that $\eta_{x}$ and $\eta_{y}$ (as well as $\phi_{x}^{0}$ and $\phi_{y}^{0}$) are very nearly degenerate in mass and can
coannihilate.  In order for coannihilation to have a significant effect on the relic density, it is necessary for some of the coannihilation
channels to dominate over the normal annihilation channels.  This happens in {\it mSUGRA} when the lightest neutralino can coannihilate with the
lightest stau~\cite{Ellis:1998kh,Gomez:1999dk} or stop~\cite{Ellis:2001nx}.  It is also possible to have coannihilation with another
neutralino~\cite{Birkedal-Hansen:2002sx} both in {\it mSUGRA} in the 'focus point' region~\cite{Feng:2000gh} and also in {\it rSUGRA}~\cite{Birkedal-Hansen:2002am} and string derived models~\cite{Birkedal-Hansen:2001is}.  In LH models, the net effect is expected to be qualitatively similar
to supersymmetric models.  Including coannihilation effects, like including additional annihilation channels, will reduce the relic density.

It is also possible that significant entropy can be generated between the freeze-out temperature of $N_{1}$ and the present temperature.  
When the relic density is calculated, the quantity that is evolved is the number density per comoving volume.  This is computed by dividing the number density by the entropy density: $Y_{N_{1}} = n_{N_{1}}/s$.  In making this redefinition, it is assumed that entropy is neither added to nor removed from the system under study.  However, such an assumption about the entropy need not hold.  If entropy is added to or taken away from the universe, then the number density per comoving volume will not obey the simple standard evolution. 
In terms of an existing relic density, the net effect will be to multiply a
particle's relic density by a factor of roughly $S_{old}/S_{new}$.  $S_{old}$ is the entropy per comoving volume before the entropy change
occurred and $S_{new}$ is the entropy per comoving volume after the change in entropy occurred.  The end result is again a reduction in the relic
density for the case of entropy generation\footnote{It is possible, however, to have both entropy production and relic density increase through the decay of a heavy modulus~\cite{Moroi:1999zb,Giudice:2000ex}.}.  In some sense this generic reduction in relic density is good because it also reduces
the amount of parameter space that is excluded on cosmological grounds  (where $\Omega_{dm} h^2 > 0.129$).  However, it also results in a reduction in the amount of cosmologically preferred parameter space (where $0.094 \lsim \Omega_{dm} h^2 \lsim 0.129$).  Thus, only the regions with $\Omega_{dm} h^2 \lsim 0.094$ are enlarged.  Such regions do not supply enough relic density from the $N_{1,2}$ particles, so one might have to appeal to dark matter candidates outside of the LH model to meet current dark matter density measurements.


\section{Conclusion}
We have investigated the possibility of having a dark matter 
candidate arise from a 'little Higgs' model.  We found regions of 
parameter space which satisfy the relic density requirement \\
$0.094 \lsim \Omega_{dm} h^2 \lsim 0.129$.  These regions require a dark matter candidate that either has a mass between roughly $m_{h^{0}}/2$ and $m_{h^{0}}$ or has a mass above $400 \GeV$.  The low-mass parameter space additionally requires the candidate to have either 'super-weak' coupling or very specific masses of $m_{h^{0}}/2 \pm 3 \GeV$ or $m_W$.  This is somewhat different from the standard dark matter scenario in theories involving
broken supersymmetry and stems from the LH dark matter candidate being a scalar and also not having any direct couplings to the neutral vector bosons.  We discussed additional possible contributions to the relic density calculation of the dark matter candidate.  We found that the additional contributions tended to result in an even lower relic abundance.  

The standard estimate for dark matter gives that weak couplings
and masses $\OO(10 \GeV) - \OO(1 \TeV)$ result in the right order of magnitude 
for the dark matter abundance.  This numerology indicates that dark matter
might be related to electroweak symmetry breaking or the physics 
stabilizing the weak scale.  In this note we have shown that it 
seems possible to accommodate sufficient dark matter  for theory
space models with exact discrete symmetries.  If this had not been
the case either by over-depletion or by a breaking of the discrete
symmetry, then there would still be the usual dark matter problem.
Through  standard calculations using partial-wave unitarity it is 
possible to put an upper limit
of $550 \TeV$ on the mass of the dark matter particle\cite{Griest:1989wd}.  
By using a more refined calculation assuming weak coupling, the limit can 
be brought down to about $1.8 \TeV$~\cite{Jungman:1995df}.
If there is insufficient dark matter from such theory space models and if the dark matter is 
related to the weak scale, then it must come from physics in the
ultraviolet and the particles must have masses roughly $\OO(f)$ or less
if they are weakly coupled.  However, it should also be noted that it is possible that dark matter can originate from axions or also a sector unconnected to
the Standard Model except through gravity.

In conclusion, we have found viable dark matter in a prototypical LH model.  Requiring the correct relic density forces the LH model to one of three regions.  First, $N_{1,2}$ can have masses of $m_{h^{0}}/2 \pm 3 \GeV$ or $m_W$, resulting in resonant and threshold annihilation, respectively.  Second, it is possible to have a 'super-weak' $N_{1,2}$ with $\cos^2 \vth \sim 1$ (where the gauge contributions dominate in the $\eta$-$\phi$ mass matrix).  Lastly, the LH
model can have large $\Omega$ plaquettes to raise $m_{N_{1,2}}$.  However, both the {\it extreme} 'super-weak' regime and {\it very} large contributions from
$\Omega$ plaquettes appear to over-close the universe.

\section*{Acknowledgments}
A. B.-H. would like to thank Maxim Perelstein for numerous fruitful and enjoyable conversations.  The work of A. B.-H. was supported in part by the DOE Contract DE-AC03-76SF00098 and in part by the NSF grant PHY-00988-40.


\end{document}